\documentclass[epj,nopacs]{svjour}
\usepackage{graphicx}
\usepackage{amsmath,amssymb}
\usepackage{alphabeta}
\renewcommand{\d}{{\rm d}}
\newcommand{\e}{{\rm e}}
\usepackage{color}

\begin{document}
\title{Domain formation and self-sustained oscillations in quantum cascade
    lasers}
\author{Tim Almqvist \inst{1}
  \and David O. Winge \inst{2}\and Emmanuel Dupont \inst{3}\and Andreas Wacker\inst{1}
}

\institute{Mathematical Physics and NanoLund, Lund University, 22100 Lund, Sweden \and 
Synchrotron Radiation Physics and NanoLund, Lund University, 22100 Lund, Sweden \and 
Advanced Electronics and Photonics Research Centre, National Research Council, Ottawa, Ontario K1A 0R6, Canada
}
\date{{\bf Accepted mansucript}, to be published in EPJ B, Topical issue: Non-Linear and Complex Dynamics in Semiconductors and Related Materials 2019}
%
\abstract{We study oscillations in  quantum cascade lasers due to
  traveling electric field domains, which are observed both in
  simulations and experiments. These oscillations occur in a range of
  negative differential resistance and we clarify the condition
  determining whether the boundary between domains of different
  electric field can become stationary.
%
} 
\maketitle
\section{Introduction}
\label{intro}

Quantum Cascade Lasers (QCLs)\cite{FaistScience1994,FaistBook2013}
have become the most important devices for radiation in the infrared
region of the optical spectrum \cite{RazeghiApplOpt2017} and are also
promising for THz applications \cite {RobenOptExpress2017}. They are
based on electronic transitions between quantized states in the
conduction band of semiconductor heterostructures, which enables a
large flexibility to define the transition energy. QCLs are pumped
electrically, where a sequence of scattering and tunneling transition
fills the upper laser state and empties the lower one. While such an
active module has a typical size of 50 nm, the repetition of a large
number of modules (the name-giving cascade) allows to fill the optical
wave\-guide with active material. Here the design is based on the
assumption, that the applied bias drops homogeneously over all
modules. However, tunneling between quantized states is prone to
specific resonances \cite{CapassoAPL1986} and the appearance of
negative differential conductivity (NDC) for fields above the resonance
point.  In contrast to the resonant-tunneling
diode \cite{ChangAPL1974}, where this property is directly reproduced
in the current-bias relation, the cascaded structure of QCLs provides
an instability in the homogeneous voltage drop. This situation is
analogous to the well-studied domain formation in Gunn
diodes \cite{GunnIBM1964,KroemerIEEETransElectronDev1966,ShawBook1992}
and  superlattices
\cite{EsakiPRL1974,BonillaPRB1994,KastrupAPL1994,WackerPhysRep2002,BonillaRepProgPhys2005}.

\begin{figure}[b]
\resizebox{\columnwidth}{!}{%
 \includegraphics{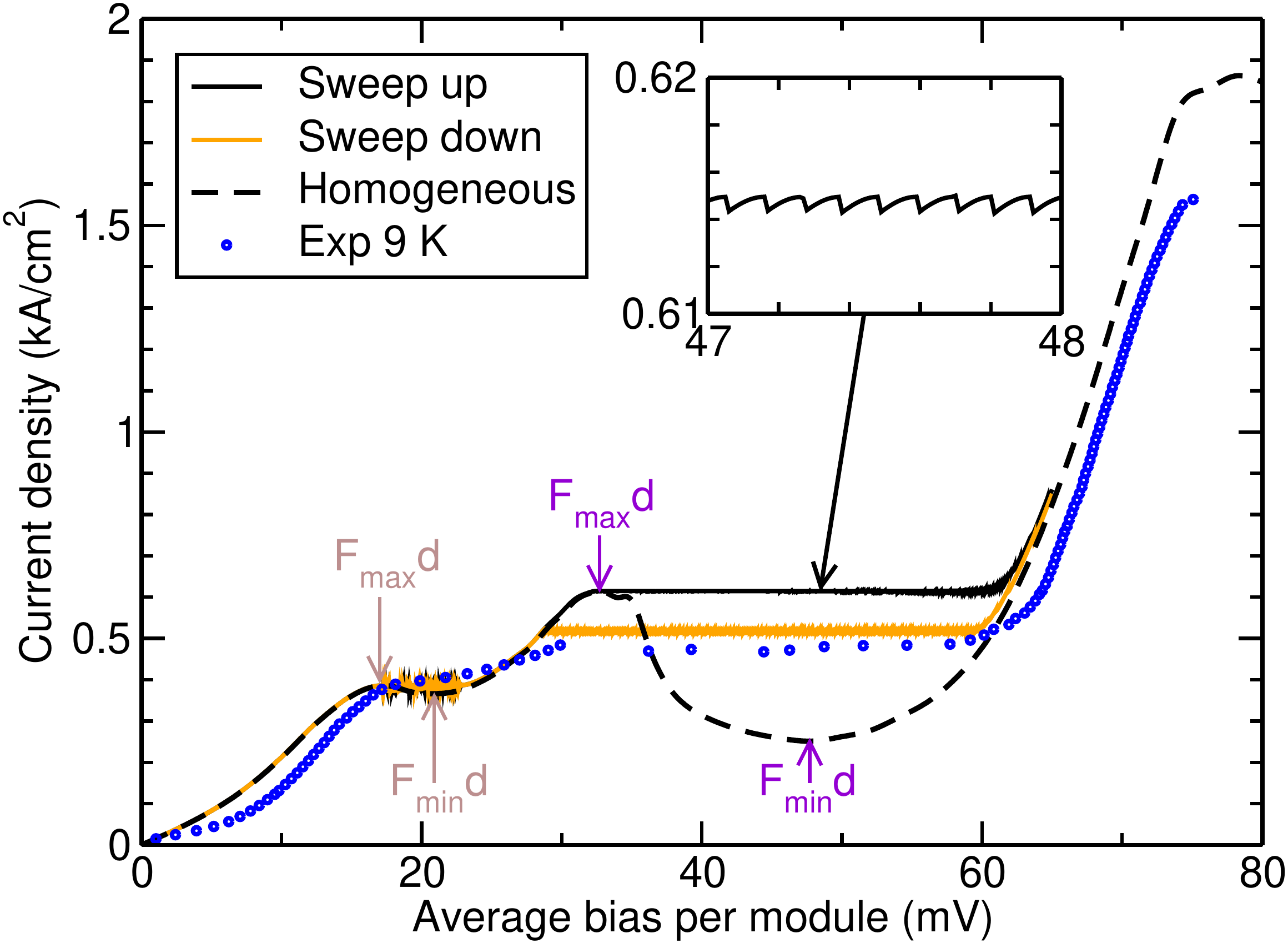}}
\caption{Simulated current-bias relation with domain formation for a
  phonon temperature of 77 K. $J_\textrm{hom}(Fd)$ (dashed line) is
  calculated by our NEGF simulation and shows two NDC regions between fields
  $F_\mathrm{max}$ and $F_\mathrm{min}$, which denote
  maxima and minima of current. The full black/orange line
  shows the result with domain formation upon sweep-up/down of the
  bias, respectively.  An enlarged section for sweep up is shown in
  the inset. The blue dots show corresponding experimental data. }
\label{FigIV}       
\end{figure}

The formation of \textit{stationary electrical field domains} in QCLs
has been observed experimentally by the characteristic saw-tooth
structure of the current-bias characteristics
\cite{LuPRB2006,WienoldJAP2011} and a spatially resolved measurement
of the bias drop over the sample \cite{DharSciRep2014}. Oscillatory
behavior has also been reported in the NDC region
\cite{SirtoriIEEEJQuant2002} which was, however, interpreted to be
determined by the circuit similar to resonant tunneling
diodes. Specifically, we consider the THz-QCL with scattering
injection discussed in \cite{DupontJAP2012}. This structure exhibits
characteristic regions of NDC in the simulated current density
$J_\textrm{hom}(Fd)$ for identical bias drop $Fd$ over all modules (so
that periodic boundary conditions apply), as shown in
Fig.~\ref{FigIV}.  
 Experimentally, current plateaus are observed
for biases and currents below threshold.
Here we show, both by experiments and simulations, that
\textit{oscillating electric field domains} occur in our device,
similar to the case of superlattices\cite{KastrupPRB1995},
and discuss the conditions for their occurrence.

\section{Simulating domain formation}
\label{SecSimulations}
\begin{figure}[t]
\begin{center}
\resizebox{\columnwidth}{!}{%
 \includegraphics{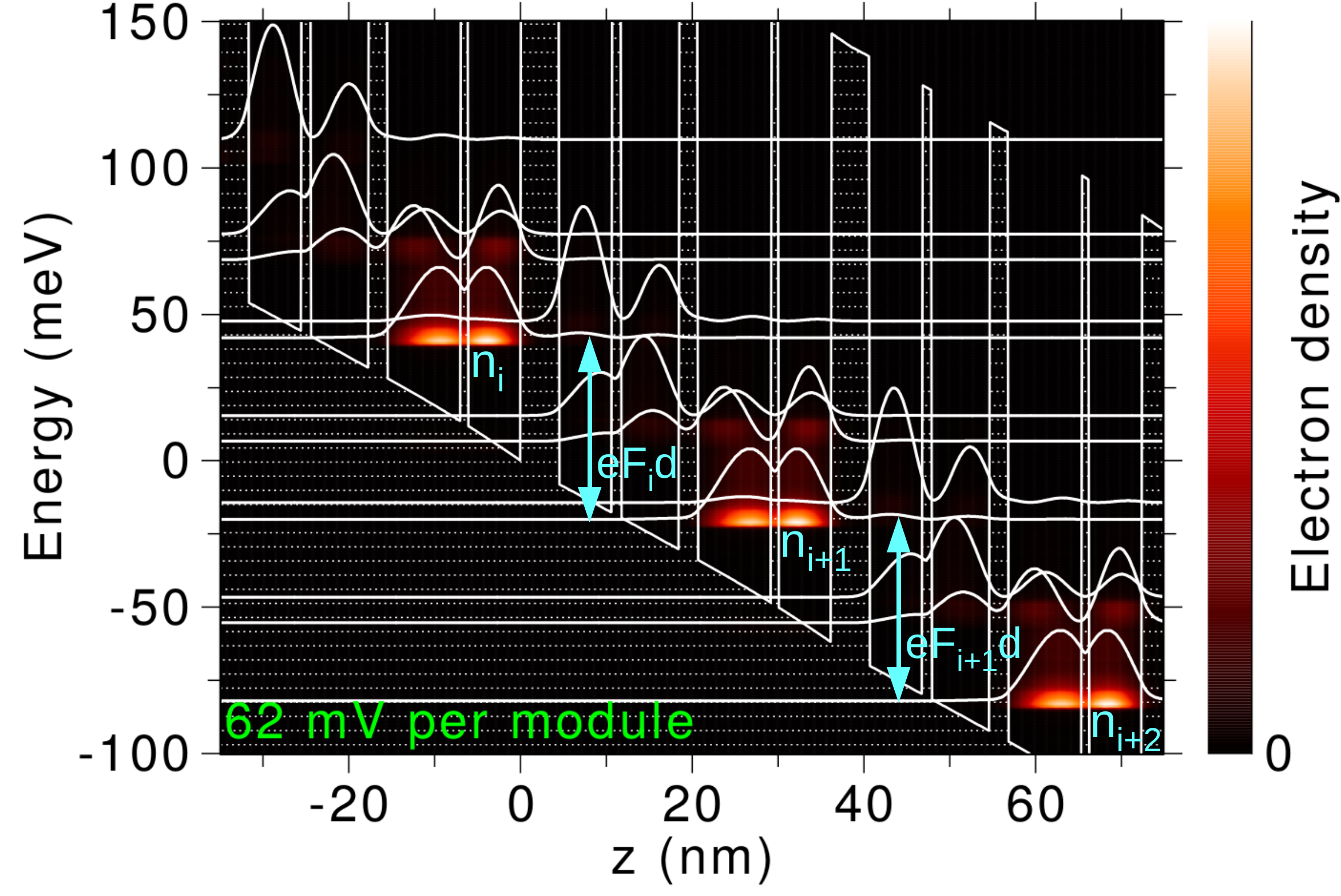}}
 \end{center}
\caption{Simulated electron density based on a homogeneous bias drop and description of the
  notation for electron sheet densities $n_i$ and potential drop $F_i d$.}
\label{FigProfile}       
\end{figure}

In order to simulate domain formation, we assume that the sheet
electron density $n_i$ of each module is essentially confined in a
narrow region before the thickest barrier. This is confirmed by our
non-equilibrium Green's function (NEGF) simulations as shown in
Fig.~\ref{FigProfile}.
(For details, see \cite{WingeJAP2016}, where all
parameters are givens. As the only
  difference we use a slightly increased rms height $\eta=3$ {\AA} for
  the interface roughness here, which appears to match better with the
  experimental results. We have no direct experimental information
  regarding the quality of interfaces.)
The same location of charges
was actually deduced experimentally \cite{DharSciRep2014}. The current
to the next module is then essentially determined by the bias drop
$F_i d$, where $F_i$ is the average electric field between the
electron densities located in module $i$ and $i+1$ and $d=36.12$ nm is
the thickness of a single module.  Following a common approach in
superlattices \cite{WackerPhysRep2002}, this current density is
approximated by
\begin{equation}
J_{i\to i+1}=
J_\textrm{hom}(F_id)\frac{n_i-n_{i+1}\e^{-eF_id/k_BT}}{N_d-N_d\e^{-eF_id/k_BT}}\, ,
\label{EqCurrentApprox}
\end{equation}
which assumes that the current density  is essentially proportional to
the electron density $n_i$ in the injecting module and is 
reduced by a backward current from  thermal excitations in the receiving module $i+1$.
This expression is normalized by the sheet doping density $N_d=3.25\times 10^{10}\textrm{cm}^{-2}$ per module, so that the
homogeneous current is recovered for $n_i=n_{i+1}=N_d$. Then the
continuity equation provides
\begin{equation}
  e\frac{\d n_i}{\d t}= J_{i-1\to i}- J_{i\to i+1} \quad\textrm{for } i=1,\ldots N
  \label{EqContinuity}
\end{equation}
where $N=276$ is the number of modules. $e$ is the positive elementary
charge and we redefine the sign of electric field and current density,
so that their signs match the forces and velocity of carriers,
respectively. In Eq.~(\ref{EqContinuity}) we need the boundary
currents, which we approximate as
\begin{equation}
 J_{0\to 1}=\sigma F_0\quad   J_{N\to N+1}=\sigma F_N \frac{n_N}{N_d}
\end{equation}
with a phenomenological boundary conductivity $\sigma$,
as justified for the Gunn diode in Ref.~\cite{KroemerIEEETransElectronDev1968}. These Ohmic boundary conditions are common for superlattices, see \cite {WackerPhysRep2002,BonillaBook2010} for a more detailed discussion.
Unless
mentioned otherwise, we use $\sigma/d=50 \textrm{ A/(cm}^2\textrm{mV)}
$, which is steeper than the slopes in $j_\textrm{hom}(Fd)$. This
mimics a good
Ohmic contact between the QCL heterostructure and
  the substrate with an adjacent highly doped ($8\times 10^{17}/\textrm{cm}^3$)
  GaAs layer implying
a small bias drop $F_0d$. For
given electron densities $n_i$, the electric fields are obtained from
the total bias $U$ dropping over the QCL 
\begin{equation}
U=\sum_{i=0}^N F_id
\end{equation}
and 
Poisson's equation
\begin{equation}
  F_{i}-F_{i-1}=\frac{e}{\epsilon_0\epsilon_r}(n_i-N_d) \quad\textrm{for } i=1,\ldots N\, .
  \label{EqPoisson}
\end{equation}
where $\epsilon_r= 13$
is the average relative dielectric constant.
For fixed bias $U$, this model has been applied for QCLs in \cite{WienoldJAP2011}. Here we consider also
the operation by a fixed bias source $U_0$ via a serial load resistance $R$. Following \cite{WackerJAP1995} we find
\begin{equation}\begin{split}
  \frac{\d U}{\d t}=& \frac {1}{C_\textrm{QCL}+C_p}\left(\frac{U_0-U}{R}-I\right)\\
  &\textrm{with }
  I=\frac{A}{N+1}\sum_{i=0}^N J_{i\to i+1}\, .
  \label{EqBiasDiff}
\end{split}\end{equation}
Here $C_\textrm{QCL}=\epsilon_r\epsilon_0A/(N+1)d$ is the geometrical capacitance of the QCL structure with $A=0.144 \textrm{ mm}^2$  and $C_p$ is an additional parallel capacitance, which we actually set to zero here.

The set of equations
(\ref{EqCurrentApprox},\ref{EqContinuity},\ref{EqPoisson}) can be
viewed as the discretization of a drift-diffusion model combined with
Poisson's equation  (as appropriate for the Gunn effect
\cite{KroemerIEEETransElectronDev1966}). Here, the drift velocity
$v(F)$ and the diffusion coefficient  $D(F)$ are given by
  \[
  v(F)=\frac{dJ_\textrm{hom}(Fd)}{eN_d}\quad \textrm{and }
  D(F)=\frac{d^2J_\textrm{hom}(Fd)}{eN_d
    \left(\e^{eFd/k_BT}-1\right)}\, ,
  \]
  respectively, which satisfy the Einstein relation in the linear
  regime for low fields. This allows for analytic insight into the
  dynamics of wave-fronts \cite{BonillaBook2010,BonillaSIAM1997}.
  However, for many features, such as the existence of stable field
  domains in a finite range of currents, the discreteness due to the
  semiconductor heterostructure is essential.

\section{Simulation results}
\label{SecSimResults}
\begin{figure}[t]
\begin{center}
\resizebox{0.8\columnwidth}{!}{%
 \includegraphics{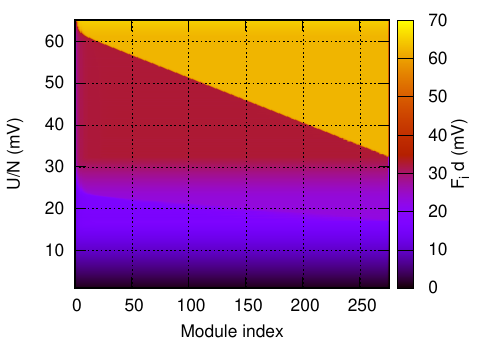}}
 \end{center}
\caption{Simulated field distribution (in color scale) as a function
  of position and total bias, obtained by sweeping up the bias. (For
    $U/N\sim 20$ mV the domain boundary oscillates in time and typical
    snapshots are displayed.)}
\label{figFieldDistr}      
\end{figure}

Figure~\ref{FigIV} shows the simulated current bias relations using a
fixed bias $U$. Upon sweeping up the bias, we find two current
plateaus; after the first current peak around an average bias of 17 mV
per module and also after the second current peak around an average
bias of 33 mV per module. The current depends on the sweep direction,
a common feature, which has been discussed for superlattices in
detail, where multi-stability was observed \cite{KastrupAPL1994}.  The
corresponding distribution of the electric field is shown in
Fig.~\ref{figFieldDistr}. Within the plateau regions, there are two
domains with specific electric fields. An increase of bias is
reflected in a shift of the domain boundary, where the high-field
domain expands.

\begin{figure}[t]
\begin{center}
\resizebox{\columnwidth}{!}{%
 \includegraphics{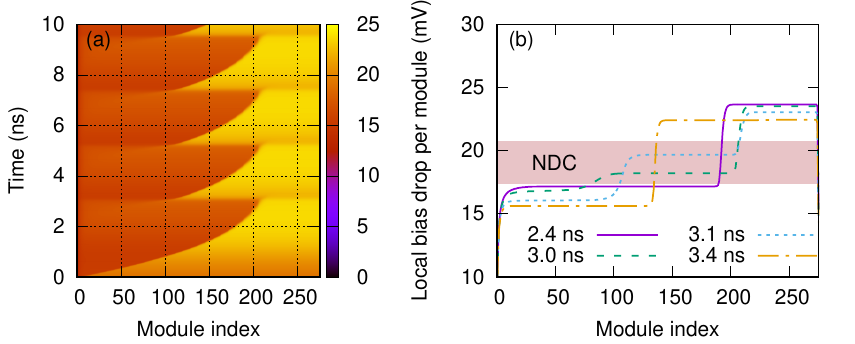}}
 \end{center}
\caption{(a) Simulated field distribution (in color scale) as a
  function of position and time for a fixed average bias $U/N=19$
  mV. Panel (b) shows the field distribution at particular times.}
\label{figOscFieldDistr}       
\end{figure}

\begin{figure}[t]
\begin{center}
\resizebox{0.8\columnwidth}{!}{%
 \includegraphics{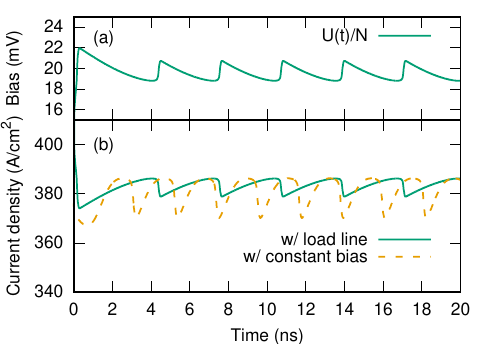}}
 \end{center}
\caption{ (a) bias and (b) current density for simulations 
  with $R=50$ \textOmega~ and $U_0=33$ V in comparison with the
  bias controlled case (dashed)
  with $U/N=19$ mV as considered in Fig.~\ref{figOscFieldDistr}.}
\label{figOscU0}       
\end{figure}

For the first plateau, the simulations provide actually oscillatory
behavior as shown in Fig.~\ref{figOscFieldDistr}. These are due to
moving domain boundaries (also referred to as fronts), where the
excess charge travels along the
direction of electron motion. For $J=300 \textrm{A/cm}^2$ we obtain an
average drift velocity of $J/(N_d e)=60$ modules/ns, which is
comparable to the average front velocity of 50 modules/ns
observed in Fig.~\ref{figOscFieldDistr}(a). During the motion of the
domain boundary, the electric fields in both domains need to increase
in order to maintain the fixed bias.
Eventually, the electric field in the low-field domain reaches
the NDC region (e.g. at $t\approx 3$ ns) and
becomes unstable. Thus, a new accumulation layer is formed within a
fraction of a nanosecond, which then starts its motion like the one
before. If the sample is operated via a load resistor, the bias $U$ is
actually oscillating as shown in Fig.~\ref{figOscU0}. In this case the
oscillation frequency is reduced.

\begin{figure}[t]
 \begin{center}
\resizebox{\columnwidth}{!}{%
 \includegraphics{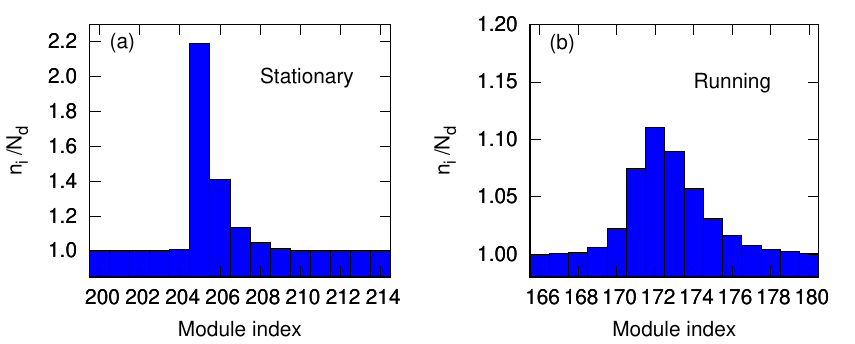}}
 \end{center}
\caption{
  Charge distribution of accumulation layer for domains at an average bias $U/N=40$ mV (left with stationary domain) and 19 mV (right with running domain)}
\label{FigAccumulation}       
\end{figure}

Such oscillations are generic for extended systems with NDC and have
been well-studied for the  Gunn Diode and superlattices
\cite{ShawBook1992,BonillaBook2010}. In continuous systems, such as
the Gunn diode, any charge
accumulation typically
travels due to the drift velocity (albeit
the actual speed may differ \cite{KroemerIEEETransElectronDev1966})
and thus stationary fronts are only possible in the vicinity of the contacts.
However, for discrete systems, such as superlattices and QCLs, these fronts
can become stationary for a finite range of currents
  \cite{BonillaPRB1994,KastrupAPL1994,LuPRB2006,CarpioPRE2000},
so that the oscillations disappear. Fig.~\ref{FigAccumulation}
shows the electron density distribution close to the domain boundary
in both plateaus. For the first plateau, see Fig.~\ref{FigAccumulation}(b),
the excess electron density
$n_\mathrm{excess}=\epsilon_0\epsilon_r(F_\mathrm{high}-F_\mathrm{low})/e$,
required to change the field between both domains, is spread out over
several modules, which resembles the continuous case of the Gunn
diode. In contrast, for the second plateau, see
Fig.~\ref{FigAccumulation}(a), a large part of the excess density is
essentially located in one module and is trapped by the heterostructure, so
that a stationary front appears \cite{WackerPRB1997a}. This situation
should be stable, if the NDC region is crossed within one module. Then at
least the part
$n_\mathrm{NDC}=\epsilon_0\epsilon_r(F_\mathrm{min}-F_\mathrm{max})/e$
of $n_\mathrm{excess}$ is located within one module, where
$F_\mathrm{max/min}$ denote the position of the current
maximum/minimum at the borders of the NDC region, respectively, as indicated in Fig.~\ref{FigIV}. In
order for this to happen, the current in a module with the excess
density must match the current $j$ of the other modules. Using the
approximation for the current (\ref{EqCurrentApprox}) under neglect of
the backward currents
[i.e. the terms with $\e^{-eF_id/k_BT}$ in Eq.~ (\ref{EqCurrentApprox})],
this implies the condition for stationary accumulation fronts
\begin{equation}
\left(\frac{n_\mathrm{NDC}}{N_d}+1\right) j_\mathrm{min}\lesssim j\, ,
\label{EqCondStationaryJ}
\end{equation}
where $j_\mathrm{min}$ is the minimum current in the NDC
region.
A slightly stronger bound has been proven in Eq.~(18) of
  \cite{CarpioPRE2000} (also including the backward currents),
  which justifies the heuristic argument given above.
As the low-field domain is only possible for $j<j_\mathrm{max}$,
where $j_\mathrm{max}$ is the peak current before the NDC region we find the 
criterion for stationary domains with accumulation regions
\begin{equation}
  \left(\frac{n_\mathrm{NDC}}{N_d}+1\right) \lesssim
  \frac{j_\mathrm{max}}{j_\mathrm{min}}\, .
\label{EqCondStationary}
\end{equation} 
A thorough proof for the stability of stationary domains under this
criterion had been provided in \cite{WackerPRB1997a} without backward current,
and follows more generally from  \cite{CarpioPRE2000}.
Based on the
homogeneous current-field relations in Fig.~\ref{FigIV}, we find
$n_\mathrm{NDC}/N_d+1=1.25$ for the first NDC region, which surpasses
the ratio between peak and valley current
$j_\mathrm{max}/j_\mathrm{min}\approx 1.05$. This  agrees with finding
only traveling fronts and oscillatory behavior in this plateau region.
For the second plateau $n_\mathrm{NDC}/N_d+1=1.92$ is lower than
$j_\mathrm{max}/j_\mathrm{min}\approx 2.44$ and stationary fronts are
possible for currents above $j=1.92 j_\mathrm{min}\approx 490 \textrm{
  A/cm}^{2}$ according to Eq.~(\ref{EqCondStationaryJ}). This value
agrees well with the minimal current density $j_\textrm{min}^\textrm{stat}=515
\textrm{ A/cm}^{2}$ for the domain states upon sweeping down the bias
in Fig.~\ref{FigIV}. In order to compare with experiment, it is
important to realize that Eq.~(\ref{EqCurrentApprox}) is an
approximation and that the minimum current  density $ j_\mathrm{min}$ is
difficult to access. Furthermore, growth imperfections can shift the
range of doping where oscillating field domains are observed in
simulations for superlattices \cite{WackerPRB1995}. However, the
condition (\ref{EqCondStationary}) should reflect the most important
trend, that stationary field domains require large doping densities
and a pronounced peak to valley ratio.

\section{Experimental results}
\label{SecExpResults}
The experimental data displayed in Fig.~\ref{FigIV} show a
clear current plateau in the second
simulated NDC region (between 34 and 60 mV), while the current is
monotonously increasing in the first simulated NDC region around 20 mV.
Using a significantly different setup, Ref.~\cite{DharSciRep2014} reported the occurrence of electric field domains in the 
second region but a homogeneous bias drop in the first region.
This indicates, that the first (weak) NDC region seen in our simulations is not visible in the device. (It also vanishes in the simulations for an increased roughness $\eta=4$ {\AA} 
or an elevated temperature of 150 K.) On the other hand, NDC and the formation of electric field domains is manifest in the pronounced second
simulated NDC region.

\begin{figure}[t]
\begin{center}
\resizebox{0.9\columnwidth}{!}{%
 \includegraphics{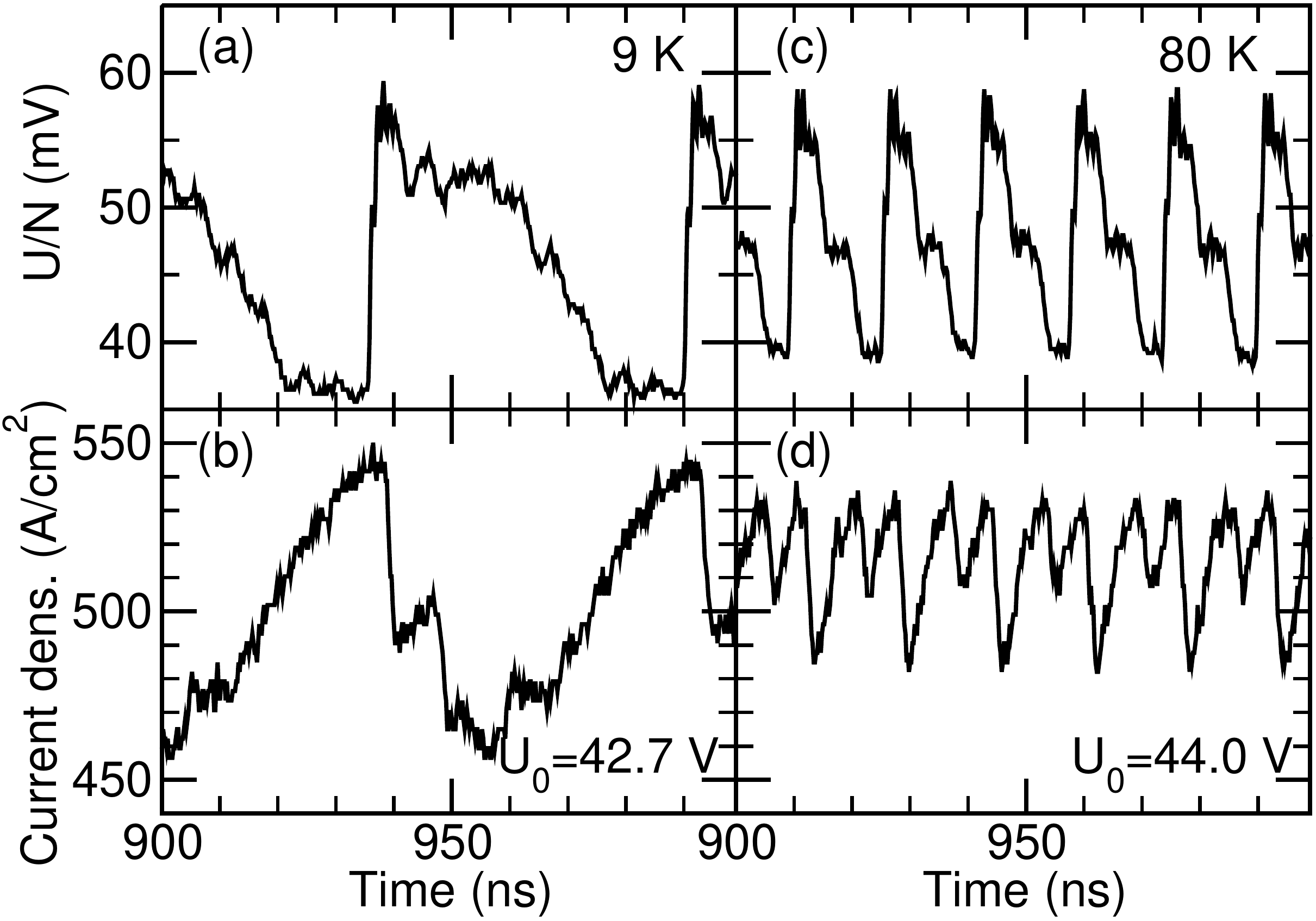}}
 \end{center}
\caption{Experimentally detected oscillations at a heat sink
  temperature of 9 K for a bias $U_0=42.7$ V [panels (a,b)] and at
  80 K for $U_0=44$ V [panels (c,d)]. The device is operated via a
  load resistance with $R=40.7$ \textOmega. Note that current and bias
  have been recorded with different cables connecting the cryostat,
  and that the frequency response of the current and voltage probes
  are different too, which all together can result to time delays of a
  few ns between the two signals.}
\label{figExpOsc}       
\end{figure}

Our measurements are based on 2 $\mu$s long pulses and the device is
operated via a load resistance of 40.7 \textOmega~ with a similar setup as
described in \cite{WingePRA2018}. In contrast to  \cite{WingePRA2018}, the laser bar studied here is indium soldered to a Si carrier with
electrodes of negligible resistivity, which simplifies the data analysis.
Using a  1~GHz bandwidth oscilloscope, we observe
oscillatory behavior in the bias as shown in
Figs.~\ref{figExpOsc}(a,c). For a heatsink temperature of 80~K,
regular oscillations are observed for external bias
$41.4 \textrm{ V} < U_0 < 48.5 \textrm{ V}$. From 48.5~V to 49.5~V
irregular oscillations are observed, in the sense that the FFT
intensity spectrum collapses. While this might indicate some chaotic scenario,
we cannot exclude, that these features are
triggered by the noise of the input pulses. Above 49.5~V,
few sporadic
short surges of voltage are still observed indicating some voltage
instabilities at the exit of the plateau. At the very beginning of the
plateau, from 41.4 to 42.1~V,
high amplitude oscillations at
fundamental frequency $\sim 27.6$ MHz are observed. At 42.1~V,
as intense but faster oscillations appear at $\sim 59$ MHz. As the
external bias further increased, the frequency of oscillations
increases slightly and its amplitude decreases monotonously.  When the
regular oscillations vanishes at 48.5~V,
the frequency has shifted to
$\sim 65.1$ MHz. The range of external voltage where the voltage
oscillations or instabilities occur is very consistent with the
observation of the current plateau. For a heatsink temperature of 9~K,
the observed frequencies are lower and the oscillations are not found
at all operation points within the plateau.

Concomitant oscillations of the total current (QCL plus voltage probe)
measured at the load resistor were also observed and are displayed in
Figs.~\ref{figExpOsc}(b,d). The current transformer used has a bandwidth
of 200 MHz, which limits the resolution here.
 
\section{Simulations with reduced contact conductivity}
\label{SecResultsReducedSigma}
The experimental observation of oscillations within the second NDC
region of Fig.~\ref{FigIV} does not agree with the simulated
stationary domains discussed in Sec.~\ref{SecSimResults}, which are
expected due to the  stability criterion (\ref{EqCondStationary}).
These stationary field domain distributions show an electron
accumulation layer separating the low-field domain close to the
injecting contact and the high-field domain at the receiving
contact.\footnote{Domain distributions with a depletion layer, which
  have the high-field domain at the injecting contact, are also
  possible. In order to be stationary, a condition similar to
  Eq.~(\ref{EqCondStationary}) exists \cite{WackerPhysRep2002}, which
  however  requires a substantially higher doping and is not satisfied
  for our device.}  Such a field distribution is only possible if the
contact field $F_0$ is small at the relevant currents, as otherwise a
high-field domain is present at the injecting contact. As discussed in
Sec.~\ref{SecSimResults}, the stationary domain states require
current-densities above $j_\textrm{min}^\textrm{stat}=515 \textrm{
  A/cm}^{2}$ as a consequence of the criterion
(\ref{EqCondStationaryJ}). In the following we use a low contact
conductivity $\sigma/d=12 \textrm{A/(cm}^2\textrm{mV)}$,
which would reflect a significant offset between the elctrochemical
  potential in the contacting layer and the levels in QCL structures.
  The chosen value of $\sigma$ implies
  $F_0d \geq 43$ mV for $j_{0\to 1} \geq j_\textrm{min}^\textrm{stat}$,
which is close to
$F_\textrm{min}d$ at the end of the NDC region. Thus the low-field
domain cannot exist at the injecting contact for $j\gtrsim
j_\textrm{min}^\textrm{stat}$ and the stable domain configurations
obtained for the second NDC region in  Sec.~\ref{SecSimResults} cannot
exist.  As a consequence we find oscillatory behavior similar to the
experimental observation.

\begin{figure}[t]
\begin{center}
\resizebox{\columnwidth}{!}{\includegraphics{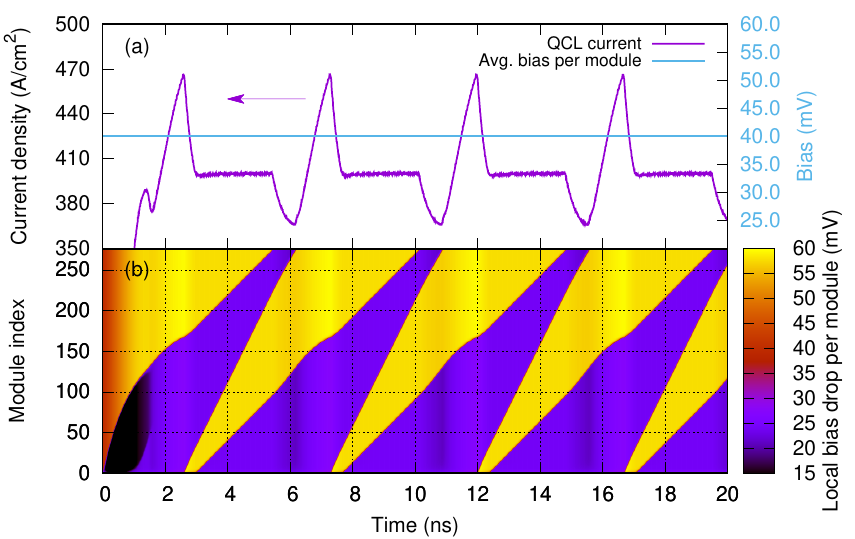}}
 \end{center}
\caption{Simulated oscillations in the second plateau for a fixed bias
  per module of 40 mV applying a reduced contact conductivity
  $\sigma/d=12 \textrm{A/(cm}^2\textrm{mV)}$.}
\label{figOsc2Plateau_constU}       
\end{figure}

Fig.~\ref{figOsc2Plateau_constU} shows oscillatory behavior for a
fixed average bias $U/N=40$ mV, i.e. in the region of the second
plateau.  Panel (b) shows a conventional electric field domain
distribution  at 11 ns. Panel (a) shows that the current density is
below $j_\textrm{min}^\textrm{stat}$  and thus the distribution is not
stationary but travels to the receiving contact. The constant bias
provides a raise in the fields and an increase in current. At 12 ns,
the field $F_0$ becomes so large that a high-field domain forms at the
injecting contact and starts traveling towards the receiving
contact. Simultaneously, the fields drop in all domains and due to the
drop of current, the injecting contact can sustain a low-field domain
again after 12.5 ns. Afterwards a characteristic period with constant
current arises, where the two accumulation fronts travel with half the
velocity of the depletion front in between. This scenario is explained
in detail for superlattice in Ref.~\cite{WackerPhysRep2002}. Around 15
ns one accumulation front and subsequently the depletion front vanish
at the receiving contact and the cycle is repeated.
  
\begin{figure}[t]
\begin{center}
\resizebox{\columnwidth}{!}{\includegraphics{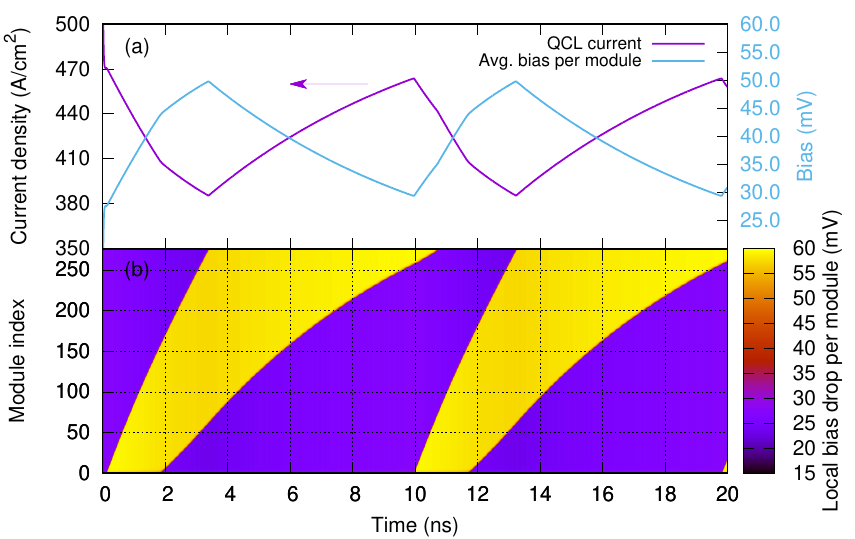}}
 \end{center}
\caption{Simulated oscillations in the second plateau for
    operating via a load resistance of 50 \textOmega~ and $U_0=41.5$ V
    applying a reduced contact conductivity $\sigma/d=12
    \textrm{A/(cm}^2\textrm{mV)}$.}
\label{figOsc2Plateau_withR}       
\end{figure}

Fig.~\ref{figOsc2Plateau_withR} shows similar domain oscillations
under circuit conditions with a load resistor. The frequency is
reduced and the signals are altered. Comparing with the electric
field distribution in panel (b), we find that the maxima and minima in
current from panel (a) are related to the creation of a depletion front at the
injection contact and its vanishing at receiving contact,
respectively. The corresponding creation and vanishing of the
accumulation front is associated with a smaller increase and decrease
of slope in the current signal, respectively. The bias behaves
precisely the opposite way. Upon varying $U_0$ we find similar results
over the entire second NDC region with average current densities
around $430 \textrm{ A/cm}^2$. This current plateau (not shown)
agrees excellently with the experimental data in Fig.~\ref{FigIV} and
the simulated oscillation frequencies are comparable  (about a factor
two larger) to the experiment (at 80 K). However, the particular
shapes of the current and bias signal differ, which might be related
to more intricate boundary currents $j_{0\to 1}(F_0)$ or to details in
the circuit such as a parallel capacitance $C_p$ not accounted for in
our simulations.

\section{Conclusion and discussion}
We demonstrated both by simulations and experimentally, that
oscillating electric field domains are possible in QCLs. Stationary
domains are favored by high doping, a large peak to valley ratio
$j_\mathrm{peak}/j_\mathrm{min}$ in the NDC region and a small excess
charge between the domains as quantified by
Eq.~(\ref{EqCondStationary}). Furthermore the injecting contact needs
to allow for the presence of a low-field domain at its vicinity for
current densities above the minimal current
$j_\textrm{min}^\textrm{stat}$ for stationary domain states.
$j_\textrm{min}^\textrm{stat}$  can be estimated by
Eq.~(\ref{EqCondStationaryJ}).
While the simulated current-bias relations agree well
with the experimental data, we did not obtain full agreement regarding
the details of the oscillations. In particular, the simulations
provide domain oscillations in the first NDC region, which appear to
be absent in the experiment. This might be related to a higher
background current in the experiment, which  counteracts the weak NDC
feature. Furthermore, the oscillation signals in the second NDC region
differ, which we can not explain now. Finally, we want to point out,
that very recently some of us observed domain oscillations in a
different QCL, which also persisted after the onset of lasing both
experimentally and by simulations \cite{WingePRA2018}.

\section{Acknowledgments}
Financial support from the Swedish Science Council (grant 2017-04287)
and Nanolund is gratefully acknowledged.

\section{Authors contributions}
Tim Almqvist did the simulations and Emmanuel Dupont the measurements
presented here. The numerical codes were developed by Tim Almqvist,
David Winge, and Andreas Wacker, who also guided the work. All the
authors were involved in the preparation of the manuscript and
approved the final version.
%

\end{document}